\def \yskip{\penalty-50\vskip3pt plus 3pt minus 2pt}
\def \reference{\par \yskip \noindent \hangindent .4in \hangafter 1}
\def \abc#1#2#3#4 {\reference#1, {\sl#2}, {\bf#3}, #4}
\def \blank {\lower 5pt\hbox to 0.75in{\hrulefill}}
\def \kms{~\rm{km}~{\rm{s}^{-1}}}

\def \AU{\rm{AU}}
\def \cm{~\rm{cm}}
\def \K{~\rm{K}}

\def \s{~\rm{s}}
\def \km{\rm{km}}

\def \erg{~\rm{erg}}

\def \yr{\rm{yr}}
\def \ves{{v_{\rm{esc}}}}

\documentstyle[12pt,aasms4]{article}

\begin{document}

\title{DETECTING PLANETS IN PLANETARY NEBULAE}
\begin{center}
Noam Soker\\
soker@physics.technion.ac.il \\
Department of Physics, University of Haifa at Oranim\\
Tivon 36006, Israel 
\end{center}

\clearpage 

\begin{center}
\bf ABSTRACT
\end{center}
 
 We  examine the possibility of detecting signatures of surviving 
Uranus-Neptune-like planets inside planetary nebulae.  
 Planets that are not too close to the stars, orbital separation
larger than $\sim 5 \AU$, are likely to survive the entire evolution
of the star. 
 As the star turns into a planetary nebula, it has a fast wind and 
a strong ionizing radiation.
The interaction of the radiation and wind with a planet may lead to 
the formation of a compact condensation or tail inside the planetary nebula,
which emits strongly in H$\alpha$, but not in [OIII].
The position of the condensation (or tail) will change over a time of 
$\sim 10$ years. 
 Such condensations might be detected with currently existing telescopes. 

{\bf Subject headings:} planetary nebulae: general 
$-$ stars: binaries: close
$-$ stars: AGB and post-AGB
$-$ stars: planetary systems

\clearpage 

\section{INTRODUCTION}

  In recent years more observational evidence and theoretical 
arguments have emerged to support the existence of extrasolar planets.  
 These include the detection of protoplanetary disks 
(e.g., O'Dell, Wen, \& Hu 1993; Strom, Strom, \& Merrill 1993) 
and Jupiter-like planets around sun-like stars
(e.g., Mayor \& Queloz 1995; 
Cochran {\it et al.} 1997), 
the theoretical argument of axi-symmetrical planetary nebulae (PNs)
formation via the influence of gas-giant planets (Soker 1996), 
the formation of comet-like objects in PNs (Dar 1998) and other 
theoretical arguments (Fargion \& Dar 1998 and references therein).  
 As the star evolves along the red giant branch (RGB) and later on the
asymptotic giant branch (AGB), it will engulf any Jupiter-like planet 
residing closer than $\sim 4 \AU$ (Soker 1996). 
 Soker (1996) argues that this results in the spinning-up of the 
star's envelope, causing it to blow axisymmetrical, rather than 
spherical, wind. 
 Planets residing further out, at orbital separation of $a \gtrsim 5 \AU$,
will survive the primary's evolution to the PN phase. 
 In this paper we examine the possibility of detecting these surviving
gas giant planets during the PN phase.
 
 That planets may reveal themselves around evolved stars was suggested 
by Struck-Marcell (1988), who proposed that SiO masers in Mira stars 
may originate in the magnetospheres of gas-giant planets. 
 This model, though, is not the popular one.  
 It requires, as noted by Struck-Marcell (1988), that several 
planets be present around many sun-like stars. 
 Dopita \& Liebert (1989; hereafter DL89) proposed that the unresolved 
compact nebula around the central star of the PN EGB 6 results from the 
ablation of a Jovian planet. 
 They further note that the ablation and formation of a compact nebula
inside a PN can be a way to detect planets. 
 Although their scenario for the formation of the compact nebula of EGB 6
is no longer accepted (Liebert, private communication),  
we follow their basic ideas here but take different approaches 
to several processes and physical parameters. 
 
 The formation of the compact tail (or nebula) is discussed in $\S 2$,
while in $\S 3$ we elaborate on some properties of the tail,
and the possibility of observing it.  

\section{THE FORMATION OF A COMPACT TAIL} 

 Two factors make the planets more likely to be detected during the PN 
phase: 
the high luminosity of the central star and its energetic wind. 
 In their model for the compact nebula of EGB 6, DL89 assume that 
the planet's distance from the central star is $2-4 \AU$. 
 Because of tidal effects planets at such close orbital separations 
will not survive the AGB phase of the primary (Soker 1996).
 We therefore take the orbital separation to be in the range of 
$a=10 \AU - 30 \AU$. 
 DL89 examined two process that ablate material from the planet.
In the first, each ionizing photon from the central star that hits 
the planet releases one atom. 
 This mechanism gives an upper limit on the planet's mass loss rate, 
since recombination must be taken into account in the dense planet's 
atmosphere. 
 DL89 then take the second approach, which they claim to be the lower limit
on the  planet's ablation rate, and take the mass loss rate to be 
somewhere between the two values. 
 In the second mechanism proposed by DL89 the gas is supplied to the 
ionization front by atoms in the planet's atmosphere having a mean free 
path about equal to the barometric scale height.   
 The mass ablation rate in the second mechanism does not depend either
on the ionization rate by the central star or on the radius of the planet.
It is proportional to the planet's mass and to one over the
square root of the planet's temperature. 

 We would like to propose a different approach to estimate the mass
ablation rate from planets. 
 Our proposed mechanism is applicable to planets having an escape velocity
of $\ves \lesssim c_s$, where $c_s = 15 (T/ 10^4 \K)^{1/2} \kms$ 
is the sound speed of the ionized planet's material. 
 McCray \& Lin (1994) study the interaction of the radiation from the
progenitor of SN 1987A with the ring around it. 
 They take the particle number ablation rate of the ring to be  
$N f$, where $N$ is the number of ionizing photons per unit time
hitting the ring, and $f=t_{\rm rec}/t_{\rm esc}$ is the ratio of the 
recombination time to the escape time of a particle from the disk. 
 We take here the same approach and assume that the planet's mass 
ablation rate is given by 
\begin{equation}
\dot M_{pr}  \simeq N \mu m_H 
\left({{\tau}\over{n}}\right)
\left({{R_p}\over{c_s}}\right)^{-1},
\end{equation}
where $\tau/n$ is the recombination time, $n$ is the total number density 
of the ablated layer, $R_p$ is the planet's radius, so that
the escape time from the planet is $R_p/c_s$. 
 $N=N_\ast (R_p/2 a)^2$ is the number of ionizing photon per unit time 
hitting the planet, where $a$ is the orbital separation,
and $N_\ast$ is the number of ionizing photons emitted per unit time
by the central star.
 For a central star temperature of $\sim 10^5 \K$ it is 
$N_\ast \simeq 10^{47} (L/5,000 L_\odot) \sec^{-1}$. 
 In equation (1) we take a conservative approach, where each ionizing photon
results in one particle, instead of two (an electron and an ion) 
as assumed by DL89.   
 Therefore $\mu=0.62$ instead of $\mu=1.3$ used by DL89.  

 We assume that the ablated mass is supplied by the hot planet's gas 
expanding radially at the sound speed, which makes sense only when 
$\ves \lesssim c_s$.
 This mass flow rate is given by 
\begin{equation}
\dot M_{pe} \simeq 2 \pi 
n \mu m_H R_p^2 c_s.
\end{equation}
We take the ablation to occur mainly in the half planet's hemisphere facing
the central star, hence the factor $2 \pi$ instead of $4 \pi$. 
 We equate the two expressions for the mass ablation rate 
and find the expression for total number density in the 
ablation layer 
\begin{equation}
n_a^2 \simeq N_\ast \tau (8 \pi R_p a^2)^{-1}.
\end{equation}
Substituting numerical values we get for total number density 
\begin{equation}
n_a \simeq 10^{10} 
\left({{L_\ast}\over{5,000 L_\odot}}\right)^{1/2}
\left({{R_p}\over{3 \times 10^9 \cm}}\right)^{-1/2}
\left({{a}\over{20 \AU}}\right)^{-1} \cm^{-3}. 
\end{equation}
 Substituting $n_a$ from equation (4) in equations (1) or (2) we find
for the mass ablation rate
\begin{equation}
\dot M_p \simeq 10^{-14}
\left({{L_\ast}\over{5,000 L_\odot}}\right)^{1/2}
\left({{R_p}\over{3 \times 10^9 \cm}}\right)^{3/2}
\left({{a}\over{20 \AU}}\right)^{-1} 
M_\odot \yr^{-1}. 
\end{equation}
 We find that each ionizing ion releases on average $\sim 0.1$ atoms. 
 This is $\sim 10 \%$ of the maximum ablation rate taken by DL89.
 The mass ablation rate is proportional to the square root of the ionizing
radiation $N$. 
This means that when the luminosity is $\sim 100$ times lower than what 
is used here, according to equation (5) each photon releases
more than one atom. 
 This is of course not the case. 
 When the ionizing radiation is too weak, equation (2) does not hold since
the radiation can't heat the ablation layer, with density $n_a$  
as given by equation (4), to $\sim 10^4 \K$.  
 The mass loss rate that can be supplied then is much below what
equations (2) or (5) give.  

 The mass ablation found here, as well as that in DL89, is applicable
to planets having escape velocities of $\ves \lesssim c_s = 15 \km \s^{-1}$.
 In the solar system it can be applicable to Uranus and Neptune, and less 
to Saturn. 
 Uranus's and Neptune's orbital separations are $\gtrsim 20 \AU$, 
and the separation will further increase as the central star loses mass
on the RGB and AGB. 
 A larger orbital separation means a lower mass ablation rate. 
On the other hand, the scale height in the atmosphere is 
$\sim R_p \ves / c_s \sim R_p$ for the planets we consider. 
 Therefore, the location of the ablation layer in the atmosphere 
is likely to be at high altitudes, which means that the effective
planet's radius is more than $R_p$.
 This increases the ablation rate. 
 
The mass first to be ablated is the mass the planets may have accreted
during the central star's AGB phase. 
 Since the Bondi-Hoyle accretion radius is $\lesssim R_p$ for low 
mass planets, the mass fraction accreted by the planets is 
$\sim (R_p/2 a)^2$, assuming spherical mass loss from the
progenitor. 
 A higher mass loss rate in the equatorial plane will increase this fraction. 
 If the total mass in the nebula is $M_n$, then the mass accreted by
the planets during the AGB phase is 
\begin{equation}
M_{\rm acc} \simeq 2.5 \times 10^{-11} 
\left( {{R_p}\over{3 \times 10^9 \cm}} \right) ^{2}
\left( {{a}\over{20 \AU}} \right) ^{-2}
\left( {{M_n}\over{M_\odot}} \right) 
M_\odot . 
\end{equation}

 The planet's magnetic field will not change much the results of this 
 section.  
 During accretion from the AGB wind, most of the
material is neutral. 
 In any case, the magnetic field can capture particles and {\it increase}
the mass accretion rate. 
 During the ablation process, at any given time a large fraction of the 
expanding planet's material is neutral, and is not influenced directly 
by the magnetic field.  
 More important, the thermal pressure of the material in the ablation layer
(with $n_a \sim 10^{10} \cm^{-3}$ and $T_a \sim 10^4 K$) is of the order 
of the magnetic pressure $P_B = 0.04 (B/ {\rm G})^2$. 
Therefore, it can expand despite the magnetic pressure and tension, 
since $P_B$ declines much faster than the thermal pressure 
as the distance from the planet's surface increases.  
 As stated earlier, due to the higher temperature of the atmosphere, 
we expect the ablation radius to be way above the planet's surface, 
where the magnetic field is even weaker. 

\section{OBSERVING THE ABLATED MATERIAL} 
 
 The ram pressure of the fast wind from the central star shapes the ablated
material near the planet into a tail.
 The total number density in the tail is estimated by equating the fast 
wind's ram pressure to the thermal pressure of the ablated gas. 
\begin{equation}
n_t \simeq 4 \times 10^8 
\left( {{a}\over{20 \AU}} \right) ^{-2}
\left( {{\dot M_{fw}}\over{10^{-7} M_\odot \yr^{-1}}} \right) 
\left( {{\dot v_{fw}}\over{1,000} \kms} \right) \cm ^{-3}.
\end{equation}
 If we assume radial expansion close to the planet's surface, 
 at a velocity of $C_s \simeq 15 \km \s^{-1}$,
then the ablated gas reaches this density at
a distance $r \sim 10^{10} \cm \sim 3 R_p$ from the planet.
 For a time $\tau$ of a few years, i.e., a period much shorter than the 
orbital period, the tail will be in the radial direction.
 Let the tail have a cylindrical shape with a radius $d$ and a length 
$L = \tau v_r$, where $v_r$ is the average radial velocity of the 
ablated material in the tail.
 We take $v_r \simeq C_s \simeq 10 \km \s^{-1}$, assuming that 
the fast wind does not accelerate it much during the short time $\tau$. 
 The total mass in the tail is $M_t = \tau \dot M p$, and its volume is
$\pi d^2 L$.
 Equating the thermal pressure in the tail to the ram pressure of the fast 
wind, i.e., taking the density from equation  (7), we find 
\begin{equation}
d \simeq 2 \times 10^{10}
\left({{\dot M_p}\over{10^{-14} M_\odot \yr^{-1}}} \right)^{1/2}  
\left({{n_t}\over{4 \times 10^{8} \cm^{-3}}}\right)^{-1} 
\cm. 
\end{equation}
 For $\tau = 3 \yr$ we find $L \simeq 10^{14} \cm \simeq 0.3 a$.  

 The real shape of the ablated tail will be different because of lateral
expansion and the Kelvin-Helmholtz instability.  
 Kelvin-Helmholtz instability modes will tear blobs from the tail
in a relatively short time (less than a year).
  These blobs will be accelerated after they are no longer behind the
planet, and in several years reach large radii in the flow.  
  Overall, we expect that the compact and dense region behind the planet
will have a characteristic volume of $\sim \pi d^2 L \sim 10^{35} \cm^3$,
and a total number density of $\sim 4 \times 10^{8} \cm^{-3}$. 
However, it will be composed of many blobs, and because of acceleration 
by the fast wind it will extend to larger distance behind the planet,
even to $\sim 10^{15} \cm$. 
 This tail will be a strong emitter of H$_\alpha$, but the density 
is too high for any significant [OIII] emission. 
  The total H$_\alpha$ emission is 
$ L_{{\rm {H}}_\alpha} \simeq 10^{27} (V/10^{35} \cm^3)
(n_t/4 \times 10^8 \cm^{-3})^2 \erg \s^{-1}$. 
 At a distance of $0.5 {\rm pc}$, the maximum angular separation between 
the star and planet will be $0.04^{\prime \prime} (a/20 \AU)$, and the 
angular separation is in the limit of currently existing telescopes.
 The tail itself will be covered by only a few pixles, depending on its 
 radial extension. 
The H$_\alpha$ flux is 
$F_{{\rm H}_\alpha} \simeq 10^{-17} \erg \cm^{-2} \s^{-1}$. 
This is not negligible, considering that the flux from 
a typical PN at such a distance is 
$\sim 10^{-11}-10^{-12} \erg \cm^{-2} \s^{-1}$,
extending over a size of $\sim 100 ^{\prime \prime}$. 
 Since the brightest region of most PNs is a shell, the intensity
in the center will be very low, facilitating the detection of the ablated
tail.  

  We conclude that such an ablated planet can be detected by 
careful analysis: (i) observation over a time $\gtrsim 1,000$ seconds,
close to the central star, by removing the central star emission; 
(ii) finding the fluxes' ratio [OIII]/H$_\alpha$, which should be $\ll 0.1$,
as opposed with the rest of the nebula for which it is $\sim 1$; 
(iii) if possible, repeating the observation after several years, looking
for a displacement of the tail (which may not be resolved and may appear as 
a condensation). 

 To give examples of candidates in a search for ablated planets, 
we list several close PNs from the list of Pottasch (1996), 
and which morphologies suggest interaction of their progenitors 
with planets (Soker 1997): 
NGC 6720 (PN G 063.1+13.9), NGC 3242 (PN G 261.0+32.0),
NGC 7293 (PN G 036.1-57.1), NGC 246 (PN G 118.8-74.7).

\acknowledgments
This research was supported by a grant from the
Israel Science Foundation.

\end{document}